\documentclass[%
 aip,
 amsmath,amssymb,
 reprint,%
]{revtex4-2}
\usepackage[utf8]{inputenc}
\UseRawInputEncoding % Very import package to fix UTF-8 related errors
\usepackage[T1]{fontenc}
\usepackage{etoolbox}
\usepackage{color}
\usepackage{dcolumn}% Align table columns on decimal point
\usepackage{graphicx}% Include figure files
\usepackage{subfigure}
\usepackage{placeins}
\usepackage{mathptmx}
\usepackage{amsmath}
\usepackage{multirow}
\usepackage{bm}% bold math

\begin{document}

% TITLE %
%============================================================================

\title{Shielded ionisation discharge (SID) probe for spatio-temporal profiling of pulsed molecular beam}

% AUTHOR %
%============================================================================

\author{Milaan Patel}
\email{milaan.patel@ipr.res.in}
\affiliation{Institute for Plasma Research, Near Bhat, Gandhinagar 382428, Gujarat, India}
\affiliation{Homi Bhabha National Institute, Training School Complex, Anushaktinagar, Mumbai 400094, India}

%second author
\author{Jinto Thomas}
\affiliation{Institute for Plasma Research, Near Bhat, Gandhinagar 382428, Gujarat, India}
\affiliation{Homi Bhabha National Institute, Training School Complex, Anushaktinagar, Mumbai 400094, India}

%third author
\author{Hem Chandra Joshi}
\affiliation{Institute for Plasma Research, Near Bhat, Gandhinagar 382428, Gujarat, India}
\affiliation{Homi Bhabha National Institute, Training School Complex, Anushaktinagar, Mumbai 400094, India}

% ABSTRACT %
%============================================================================

%\date{\today}
\begin{abstract}
In this work we report a shielded ionization discharge (SID) probe which is conceptualized, designed and implemented for measuring temporal and spatial density profiles in a pulsed molecular beam. The probe detects and provides profiles of neutrals of the molecular beam using small discharge which is assisted by the thermionic emission of electrons from a hot filament. In this article, design and characterisation of the developed probe are discussed. The performance of the probe is demonstrated by measuring the spatio-temporal profile of a 1.5 ms pulsed supersonic molecular beam. The suggested SID probe can be used to characterize and optimise the pulsed molecular beam source used in tokamak fueling, SMBI plasma diagnostics, ion beam profile monitors, cluster beam experiments, chemical kinetics, and other supersoinc beam related applications.
\end{abstract}

% DOCUMENT %
%============================================================================

\maketitle
\section{Introduction}\label{sec:intro}

Supersonic molecular beam sources are used in tokamaks for fuelling \cite{EAST_2018,NSTX_2019} and edge plasma diagnostics \cite{TEXTOR_2012,TJII_2001}, laser cluster experiments \cite{Ieshkin_2022}, charged particle beam profile monitors \cite{CERN_2022}, study of reaction kinetics and surface property studies \cite{gsi_1,gee2022,gsi_book}, etc. In most of these applications, higher beam density is desirable. As a result, over the decades, molecular beam sources evolved from effusive sources to directed beam sources \cite{nozzle_beam_1,nozzle_beam_2,abm1988} and, recently as pulsed beam sources \cite{even2014}. 

Successful implementation of molecular beam sources relies on beam quality, which is quantified in terms of its lateral density profile (or divergence) and velocity. Existing molecular beam diagnostic procedures are well established for continuous beam sources \cite{abm1988} but not for pulsed sources. Although velocity measurement methods, such as time of flight \cite{SMBIpreciseTOF, skimmer_laser_photoionization} and laser velocimetry \cite{Yan21}, have been modified for pulsed sources \cite{Christen2011, Mil2021}, continuous beam profiling methods, e.g widely used ionisation gauge based method \cite{Namba_2006, LHC2023} and its derivatives \cite{SMBI_pitot, bei1981} are yet to be realised for pulsed sources. Beam profiling is also necessary for validation and optimization of pulsed beams, which is a fundamental requirement in some applications like ion beam profile monitors \cite{CERN_2022} and surface kinetics studies \cite{gee2022}. Barring these, a robust technique to measure the time evolution of density profile of the pulsed molecular beam source is not realized. Naturally, this necessitates the importance of development of instruments or techniques to measure the temporal and spatial profile of pulsed beams.

In this work we report a small and robust probe, which is basically designed to measure the cross-sectional profile of the beam. The probe works by skimming a small percentage of the beam using a gas shield, then ionising it with a tiny discharge assisted by a thermionically generated electron from a hot filament accelerated in strong electric field. The discharge current gives the estimate of density of skimmed portion of the beam, which is equal to the local density of the beam. The density profile of the beam can be generated by scanning the probe across the beam. The temporal resolution of the probe is decided by the discharge time scale which is of the order of few $\mu s$. As a result, the probe can measure the beam profile with pulse width as short as $100 \mu s$. Using this probe, we generated temporal evolution of the cross-sectional profile of 1.5 ms pulsed molecular beam of Argon. The probe is capable of measuring beams with much shorter pulse widths. However, our measurements have constraints only from the source itself and not due to the probe. Additionally, because of its small size, it can be easily deployed as an in-situ beam calibration device.

The manuscript is organised as follows. Section \ref{sec:exp} gives information on the supersonic molecular beam experiment system and discusses the design and working of the probe. Section \ref{sec:char} describes probe characteristics, section \ref{sec:cal} discusses the calibration method, section \ref{sec:res} shows beam profile measurements and finally Section \ref{sec:con} concludes the work.

\section{Experiment setup to measure beam profile}\label{sec:exp}

\begin{figure*}[ht]
\centering
\includegraphics[width=2\columnwidth, trim=0cm 0cm 0cm 0cm, clip=true,angle=0]{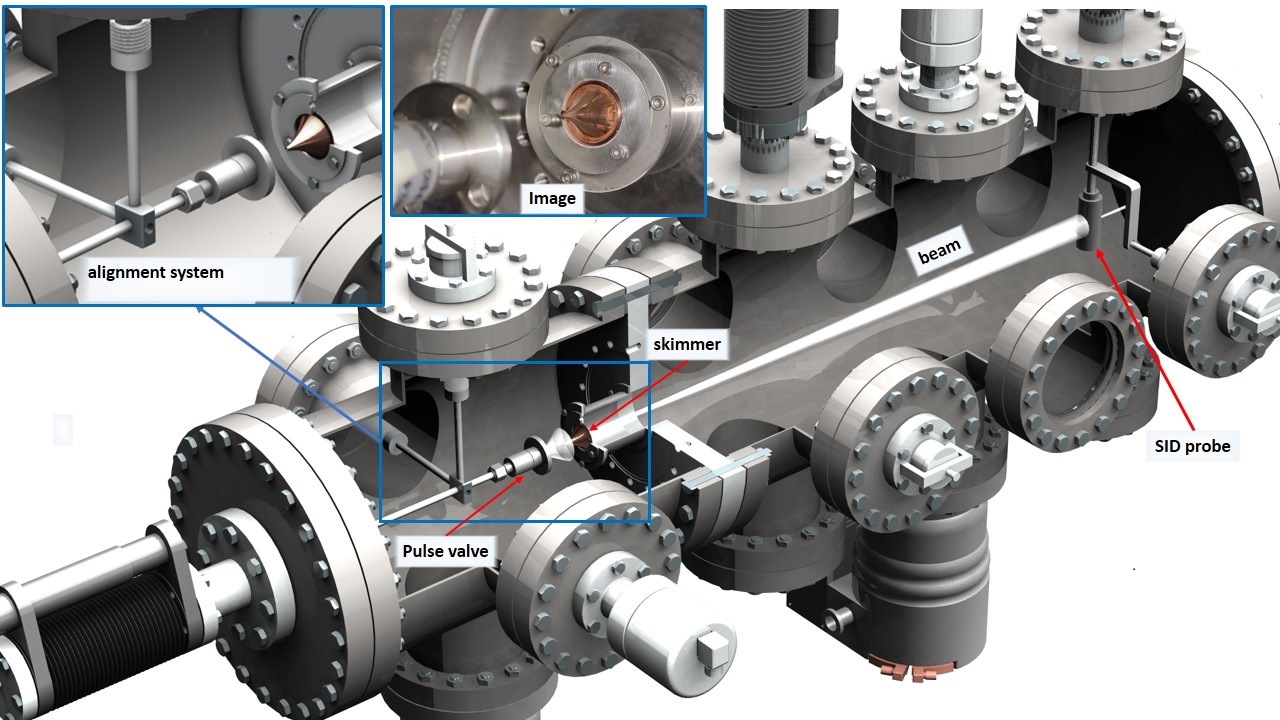}
\caption{\label{fig:expsetup} Schematic view of the supersonic molecular beam experiment system. Insets shows the zoomed model of the nozzle-skimmer alignment assembly and the photograph next to it shows the actual experiment setup.}
\end{figure*}

Typical implementation of SID probe in a supersonic molecular beam injection (SMBI) system is shown in figure \ref{fig:expsetup}. SMBI system is a prototype, developed for edge plasma diagnostics for tokamak SST-1. It has a pulsed supersonic jet source and a skimmer to extract the supersonic flow from the para-axial region of the jet into a separate chamber. The extracted flow continues as a supersonic molecular beam. SID probe is developed as a beam diagnostics and optimization tool for the SMBI system. The SMBI system and construction of the probe is described in details in the following subsections.

\subsection{Supersonic molecular beam injection (SMBI) system}

The prototype SMBI system is designed to study molecular beam dynamics and to optimize the beam for high axial density and narrow beam profile. It consists of jet and beam chambers separated by a flange containing the skimmer (skimmer-flange). 
Jet chamber consists of a Parker series 9 pulse valve with 0.5 mm orifice as source of pulsed supersonic jet. Pulse valve is mounted on a 1/4 inch SS tube which connects outside of the vessel through a gas feed-through on a bellow type translation stage. The translation stage allows the adjustment of the nozzle-skimmer distance ($Z_{NS}$) to optimise the beam. A sliding metallic sleeve is inserted on the SS tube of the pulse valve, which is coupled to two separate linear translation stages. These are used to align the pulse valve with the skimmer. 
The skimmer is made of hardened copper with orifice size of 0.4 mm, inlet half cone angle $25^o$ and edge sharpness $<10\mu m$, manufactured by beam dynamics. Skimmer is installed on a skimmer-ring which is elevated from the skimmer-flange using a pedestal to avoid the influence of background gas penetration in the free jet. Skimmer-ring is designed with slip gap to accommodate for the thermal expansion during the baking of the vessel which allows to operate the beam by avoiding misalignment even during baking at temperature of $100^oC$. The beam chamber has multiple DN 100 CF ports for installing different beam diagnostic systems that are currently in use and are planned in the future. The jet and beam chambers are pumped separately by a Pfieffer HiPace 300 turbo-molecular pump (TMP, 260$l/s$) mounted on a DN 100 CF flange and are backed by a common BOC Edwards XDS-10 scroll pump (160 lpm). Pressure inside the chambers is monitored by Pfieffer PKR251 combined full range gauges. The ultimate pressures (with baking) in the jet and beam chambers in the absence of the beam are $3\times 10^{-7}$ and $5\times 10^{-7}$ mbar respectively. 

Molecular beam of argon is generated from a pulse valve at 0.33 Hz with a reservoir pressure ($P 0$) of 5 bar and a pulse duration of 1.5 ms. Maximum dynamic pressures in the jet and beam chambers during operation are $1\times 10^{-4}$ and $1 \times 10^{-5}$ mbar, respectively. The jet and the molecular beam are illustrated in figure\ref{fig:expsetup} as transparent white volume. SID probe is mounted on a liner translation stage via an electrical feed-through. This feed-through is inserted through a hole on an L-shaped arm that is linked to another bellow type linear stage located perpendicular to the probe-feed through. The combination allows the probe to be positioned in the cross-sectional plane of the beam.

\subsection{SID Probe: design and operation}

\begin{figure}[ht]
\centering
\includegraphics[width=1.0\columnwidth, trim=0cm 0cm 0cm 0cm, clip=true,angle=0]{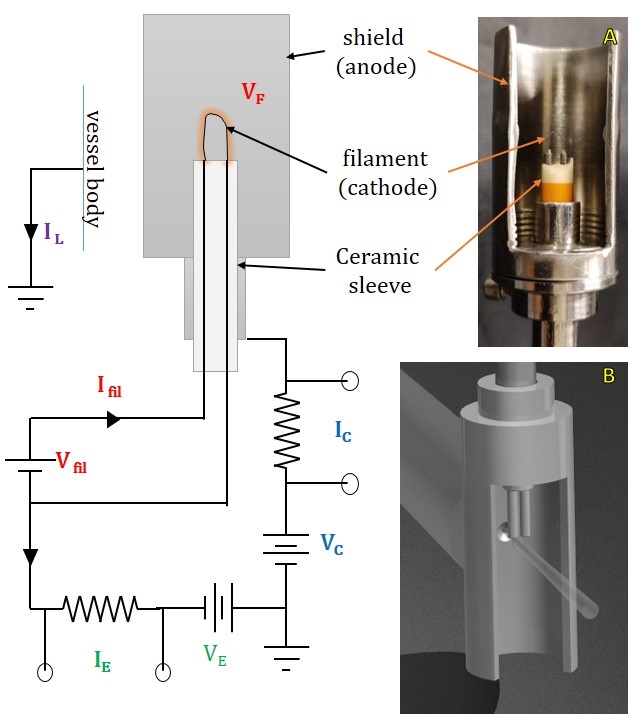}
\caption{\label{fig:probe} Schematic of SID probe with circuit diagram. Inset-A shows the picture of the actual probe and inset-B depicts a portion of the skimmed by the shield hole that creates the discharge.}
\end{figure}

The schematic of the shielded ionisation discharge (SID) probe is shown in Figure-\ref{fig:probe}. It comprises of 0.1 mm diameter filament surrounded by a 2 cm diameter cylindrical shield. The purpose of the shield is to isolate part of the beam for localised measurements, thus, preventing the remaining beam to enter the measurement region. It has a 1.28 mm hole facing the incoming molecular beam that allows limited fraction of the molecular beam to enter the probe. This skimmed fraction of the beam goes through the filament and exits the probe through rear opening of the cylindrical envelope. As the skimmed beam goes through the probe unperturbed, the detected signal can be correlated to the absolute density of the beam. Inside the probe, the skimmed beam is partially ionized by a filamentary discharge and the discharge current is measured. Value of the discharge current is used estimate the beam density using a correlation between the discharge current and neutral density established by calibrating the probe at different background pressures in the absence of the beam. Spatial resolution of the probe is determined by the size of the skimming hole on the shield. It can be increased by decreasing the size of the hole, however, at the cost of signal strength.

The filament is biased negatively, which acts as cathode, while the shield is biased positively thus, acting as anode. The filament emits electrons by thermionic emission, which move radially towards the shield because of the field generated due to potential difference $V_F$ between the anode and the cathode. The emitted electrons, thus, constitute the emission current $I_E$. The shield collects the majority of the released electrons, resulting in collected current $I_C$. A fraction of the electrons escape through the shield's rear opening to grounded body of the vacuum chamber, resulting in loss current $I_L=I_E-I_C$. Electron loss to the chamber can be controlled by biasing the shield $V_ C$ and filament $V_E$ relative to the chamber as will be explained latter in section \ref{sec:char}. If $I_L = 0$, the currents $I_E$ and $I_C$ can be referred as a discharge current. 

When the molecular beam enters the probe through the hole in the shield, it forms the region of high density along its path. This path is a projected area of the hole, having the divergence equal to that of the beam as shown in figure \ref{fig:probe}(B). High neutral density (relative to surroundings) of the beam creates favourable condition for the discharge, along the path between the hole (anode) and the filament (cathode) in presence of thermionic electrons. As a result, the discharge occurs only along the beam path between electrodes which is a cylindrical volume of length 1 cm and diameter 1.28 mm. It is absent elsewhere troughout the probe. Given the small separation between anode-cathode and low density of the beam, thermionic electrons from hot filament is necessary to strike a discharge at relatively low voltages (40-150 V). Since, the discharge occupies a small volume fraction of the probe, the discharge current $\Delta I_C$ is a tiny fraction of $I_C$ already present due to thermionic emission in the probe. Typically, the value of $I_C$ is in mA range, while the value of $\Delta I_C$ is only few $\mu A$. Once discharge is struck, it can be treated as filament discharge. The fraction of ionization is unknown, however, the current $\Delta I_C$ depends linearly on the neutral density of the beam. This linear relationship is exploited to develop a relation between the $\Delta I_C$ and beam density as discussed latter in section \ref{sec:cal}. Spatio-temporal profile of the beam is generated by scanning along the beam cross-section in steps of 2 mm.

\section{Probe charcteristics}\label{sec:char}

Values of $I_E$ and $I_C$ of the probe depend on two parameters i.e. i) filament heating current ($I_{fil}$) which is correlated with the emitted electrons, which in turn shows the electrons available to constitute discharge current and ii) strength of biasing field between anode and cathode ($V_F$) which determines the energy of electron available to initiate and sustain the discharge. Selection of $I_{fil}$ and $V_F$ directly affect the values of $I_E$ and $I_C$, hence, these are the driving parameters to be characterized. Additionally, the discharge occurs for short duration only when the beam is present. In the rest of time, the probe operates in a non-discharge regime. This makes it necessary to characterize the probe under both operating conditions. This is done by evaluating the values of $I_E$ and $I_C$ due to the variation of $I_{fil}$ and $V_F$ in both discharge or non-discharge regimes.

\subsection{$I_E$ and $I_C$ characteristics}

In non-discharge regime, emission current from the filament is a function of temperature given by Richardson Laue Dushman law \cite{Richardson, Dushman}. 
\begin{eqnarray}
I_E = S A_R T^2 e^{\frac{-\Phi}{K_BT}} 
\label{eq:RDE}
\end{eqnarray}
where $S$ is the emitting surface area, $A_R$ is the Richardson constant, $\Phi$ is the work function of the filament material and $T$ is the filament temperature. In our work, due to small size and U-bend geometry of the filament, accurate values of temperature could not be estimated, hence we only discuss it qualitatively. 
Since, temperature depends nearly linearly on electric current \cite{TI_char}, the emission current $I_E$ follows the same trend with the current $I_{fil}$ (filament current) as it does with the temperature. Hence $I_E$ vs $I_{fil}$ plots can be explained using Richardson equation. As per eq-\eqref{eq:RDE}, $I_E$ increases indefinitely with $I_{fil}$. However, in reality, $I_E$ saturates after certain rise due to space charge effect which is explained and quantified by Child Langmuir law \cite{Child, Langmuir}. Electrons emitted from the filament surround the filament forming a negatively charged region known as space charge region. This creates an electric field that prevents the filament from emitting more electrons which causes $I_E$ to saturate. However, the space charge can be swept away by applying a positive potential close to the filament, which allows more emission thereby increasing $I_E$ until fresh saturation is achieved. Maximum limit of $I_E$ is a function of sweeping field potential $V_F$ and electrode separation $d$ given by Child Langmuir law.
\begin{eqnarray}
I_{E(limit)}= \frac{4\epsilon_0}{9} \sqrt{\frac{2e}{m_e}} \frac{V_F^{3/2}}{d^2} 
\label{eq:CLE}
\end{eqnarray}
where $m_e$ and $e$ are the mass and charge of electron respectively and $\epsilon_0$ is the permittivity of space around the filament. In our probe the emission current under the biasing field increases with filament current according to Richardson Law and saturates for limiting value given by Child-Langmuir law. Hence, we define the emission behaviour in this region as "Richardson regime". At relatively higher background pressures, thermionic electrons emitted from the filament compensates for the Townsend's secondary emission electrons. This creates a discharge inside the probe at pressures higher than Richardson's region but lower than pressure required for Townsend Discharge. Depending on the excess electrons available, the discharge can occur in the pressure range $10^{-4}$ to $10^{-1}$ mbar. The emission and collection current profiles changes during discharge, therefore we refer to this as an operation in the "discharge regime." 

\begin{figure}[ht]
\centering
\subfigure[Anode \& cathode at equal opposite potential]{\includegraphics[width=1\columnwidth, trim=0cm 0cm 0cm 0cm, clip=true,angle=0]{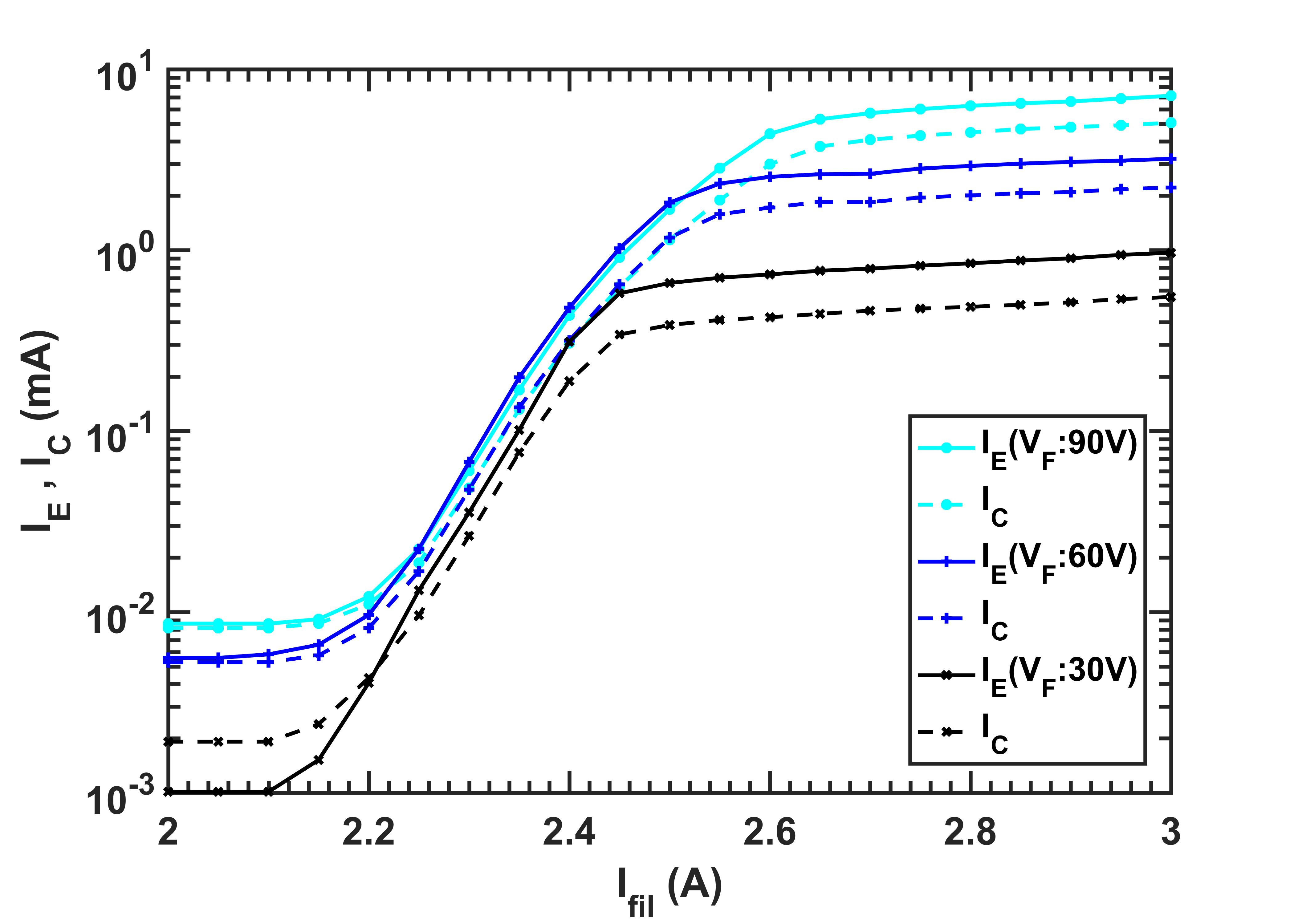}}
\subfigure[Cathode at Ground potential]
{\includegraphics[width=1\columnwidth, trim=0cm 0cm 0cm 0cm, clip=true,angle=0]{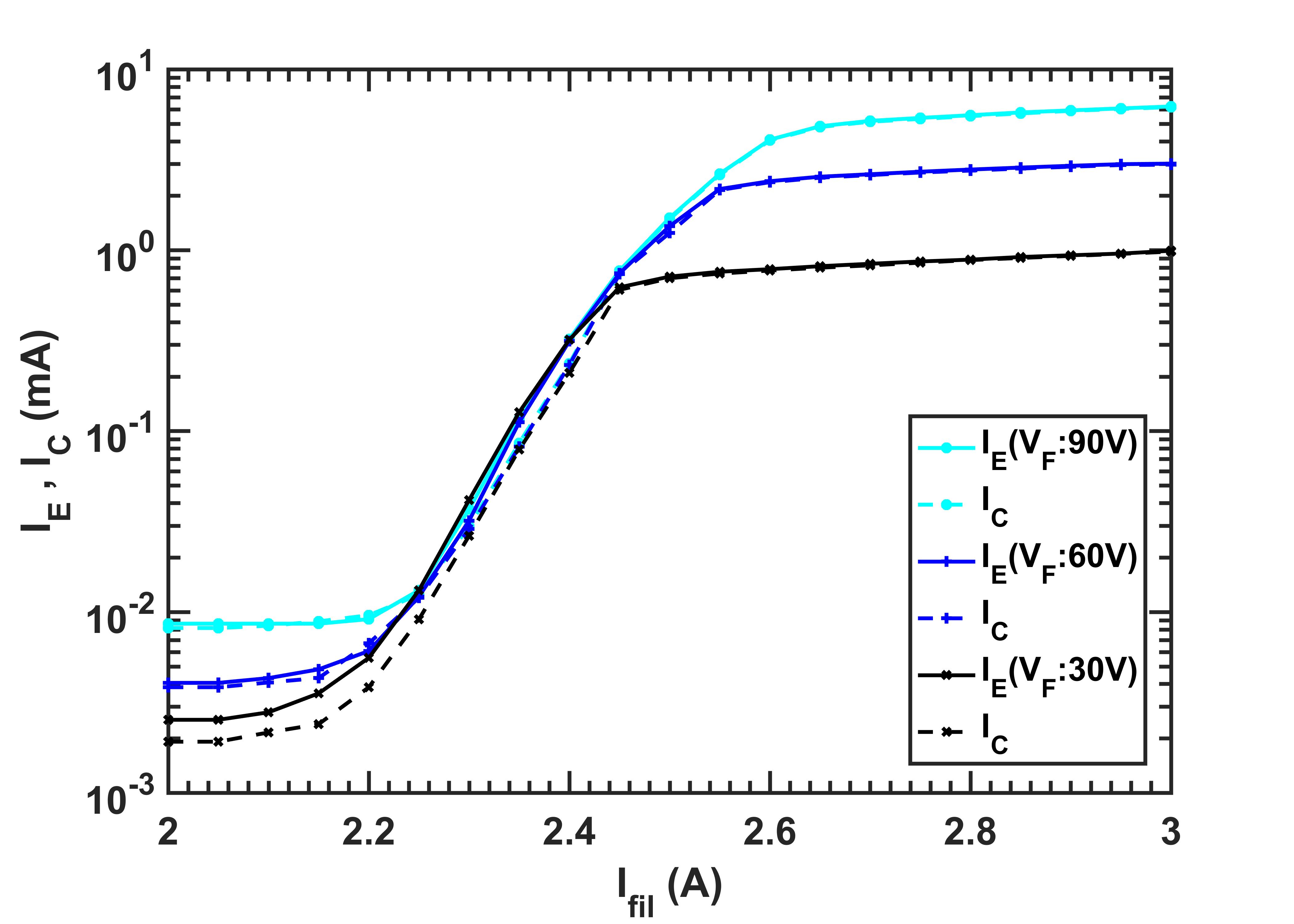}}
\caption{\label{fig:diff_I-fil} Variation of emission ($I_E$) and collection ($I_C$) current for different heating currents ($I_{fil}$) for three different field potentials ($V_F$).}
\end{figure}

Figure \ref{fig:diff_I-fil} shows the variation of $I_E$ (solid lines) and $I_C$ (dashed lines) at three different field potential $V_F$ carried out at background pressure of $2.8\times10^{-6}$ mbar. To identify modes to maximize $I_C$, measurements were carried out by biasing electrodes in two modes. In mode (a), both anode and cathode (filament) are biased with equal and opposite potentials. In mode (b), anode is biased more positively and cathode is biased close to the ground potential with a slight negative offset (5 V) to avoid fluctuations in $I_E$. It can be seen that in both the modes $I_E$ increases with $I_{fil}$ following the Richardson law until it saturates due to space charge effect. Increase in the biasing potential sweeps the space charge allowing for more emission until a fresh balance is established between the emission and the sweeping rate. $I_C$ follows the same trend and maintains a consistent difference with $I_E$ which shows that at particular $V_F$, space charge does not affect the spatial distribution of electron emission around the filament. Additionally, for both the modes, $I_E$ remains the same, which shows that emission current is only affected by relative potential difference between the electrodes and does not depend on the individual potentials of the electrodes. The difference between modes (a) and (b) is observed only in the collection effectiveness of the anode. In mode(b), $I_E =I_C$, which indicates that all emitted electrons are collected by the anode, thus minimizing the electron loss to the vessel body. Hence, during operation, the probe is biased in this mode. 

\begin{figure}[ht]
\centering
\includegraphics[width=1\columnwidth, trim=0cm 0cm 0cm 0cm, clip=true,angle=0]{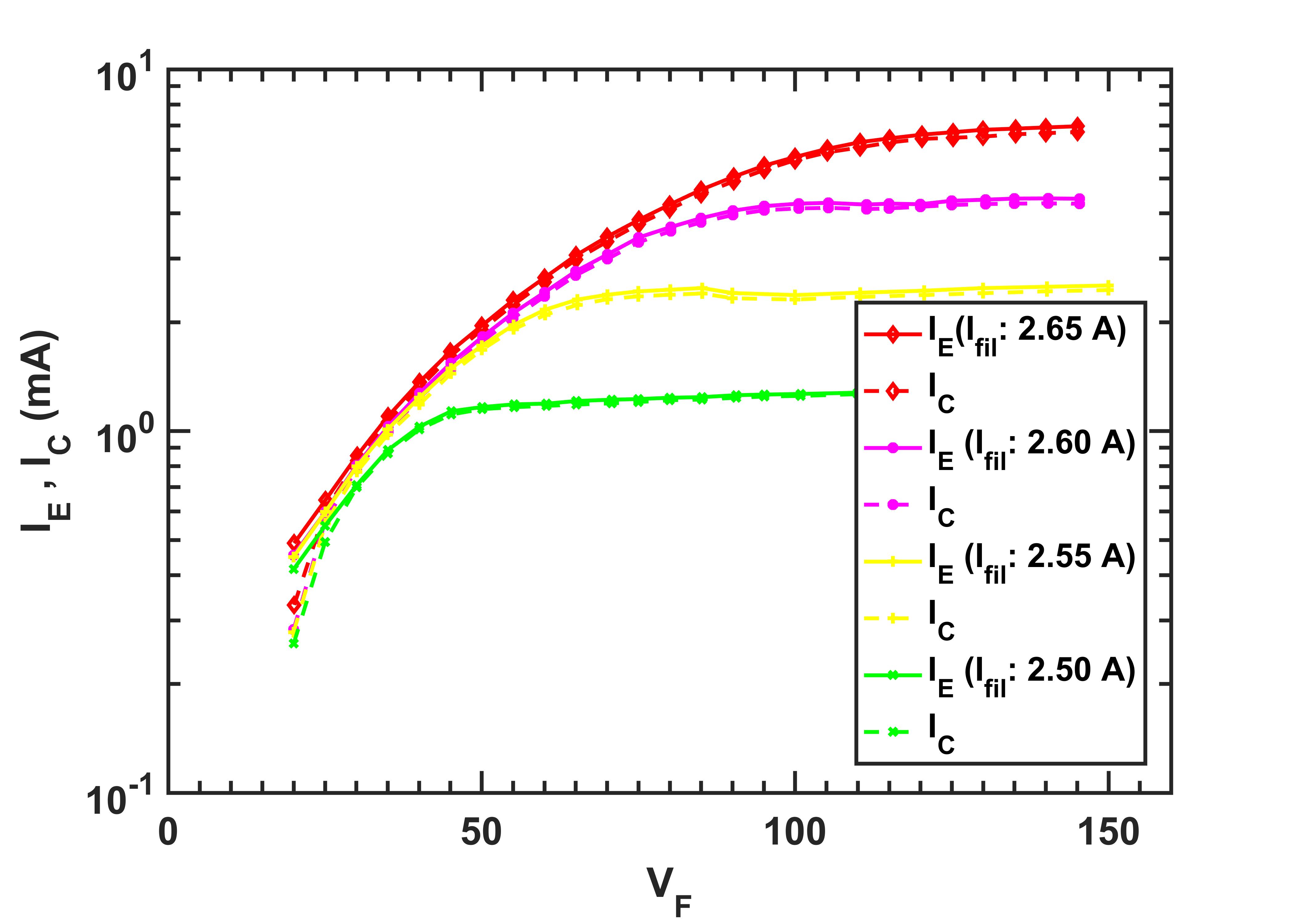}
\caption{\label{fig:diff_V-F_equal} Emission ($I_E$) and collection ($I_C$) current for different field potentials ($V_F$) for four different heating currents ($I_{fil}$)}
\end{figure}

\begin{figure}[ht]
\centering
\includegraphics[width=1\columnwidth, trim=0cm 0cm 0cm 0cm, clip=true,angle=0]{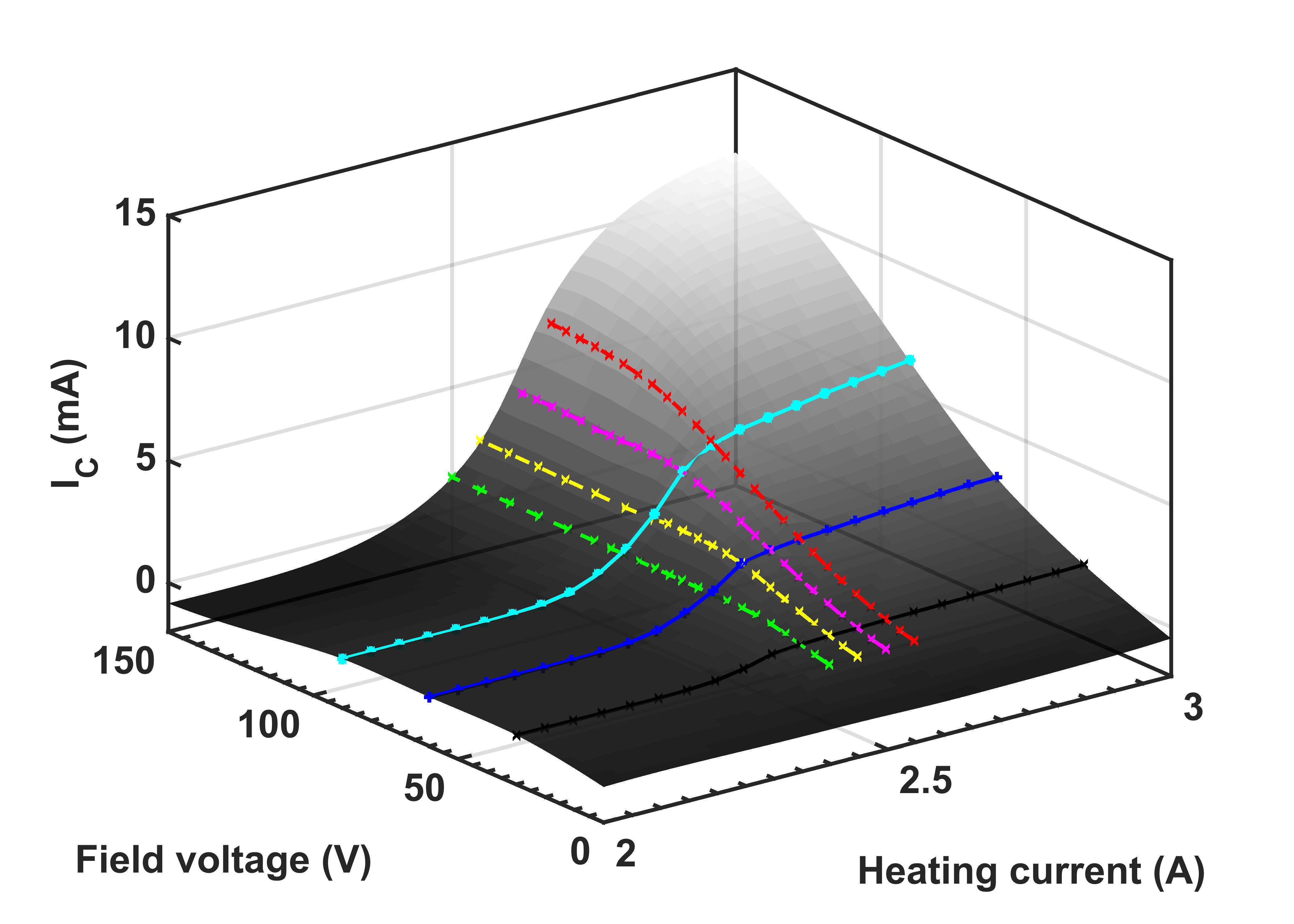}
\caption{\label{fig:richardson_characteristics} $I_C$ characteristics of SID-probe operating in Richardson regime at background pressure of $2.8\times10^{-6}$ mbar. color code of the data lines is the same as that in fig.\ref{fig:diff_I-fil}(b) and fig.\ref{fig:diff_V-F_equal}. The surface is a polynomial fit for the data.}
\end{figure}

Figure \ref{fig:diff_V-F_equal} shows the variation of $I_E$ and $I_C$ with field potential $V_F$ for different heating currents in bias mode (b). It can be seen that $I_E$ increases with $V_F$ only upto certain extent. When biasing field is strong enough to sweep all the emitted electrons, $I_E$ is limited by the filament temperature. Hence, increasing $V_F$ cannot increase $I_E$ beyond electron emission limit of the filament material for a particular temperature. This puts an upper limit on the biasing field, beyond which, the signal strength cannot be enhanced. The combined effect of $I_{fil}$ and $V_F$ can be represented as a 3D surface as shown in figure-\ref{fig:richardson_characteristics} which we term as "characteristics surface" of SID-probe for a particular background pressure. The surface representing $I_E$ can be termed as "$I_E$ characteristics" and that representing $I_C$ as "$I_C$ characteristics". Note that, figure \ref{fig:richardson_characteristics} shows characteristics surface for mode (b) where $I_E=I_C$, hence, only $I_C$ characteristics are shown.

Shape of the characteristic surfaces changes significantly once the discharge occurs, because, space charge behaves differently due to the collective behaviour of the discharge. We would like to point out that in this work, the approach used to measure density only requires the information on the neutral density during the breakdown, hence, absolute values of $I_E$ and $I_C$ are not considered. Nevertheless, it is advantageous to study it for better understanding of the working of the probe. Discharge characteristics are studied by creating the discharge in the entire volume of the probe by increasing the background pressure. Thus the neutral density during discharge is always known which is helpful in calibrating $I_C$ or $I_E$ with neutral density. Starting from a low value, the background pressure is gradually increased with the filament emitting until discharge occurs between the anode and the cathode. In a bias mode, during the operating condition of the probe (cathode close to ground and anode biased positively), it is easier to cause a discharge between the anode and the vessel than between anode and cathode due to unique characteristics of anode-cathode and their proximity to the vessel. This would create a discharge outside the probe rather than inside, which is undesirable. To avoid this, anode shield is biased close to the ground(vessel) potential and cathode filament is biased negatively. This ensures discharge occurs inside the probe, but at the cost of loss of electrons to the vessel ($I_E \neq I_C$). Hence, $I_E$ and $I_C$ characteristics need to be measured individually. As shown earlier (figure\ref{fig:diff_I-fil}), the behaviour of the probe only depends on the relative potential difference between electrodes. Hence, despite of different biasing than during operation, functionality remains unchanged.

\begin{figure}[ht]
\centering
\subfigure[emission characteristics]
{\includegraphics[width=1\columnwidth, trim=0cm 0cm 0cm 0cm, clip=true,angle=0]{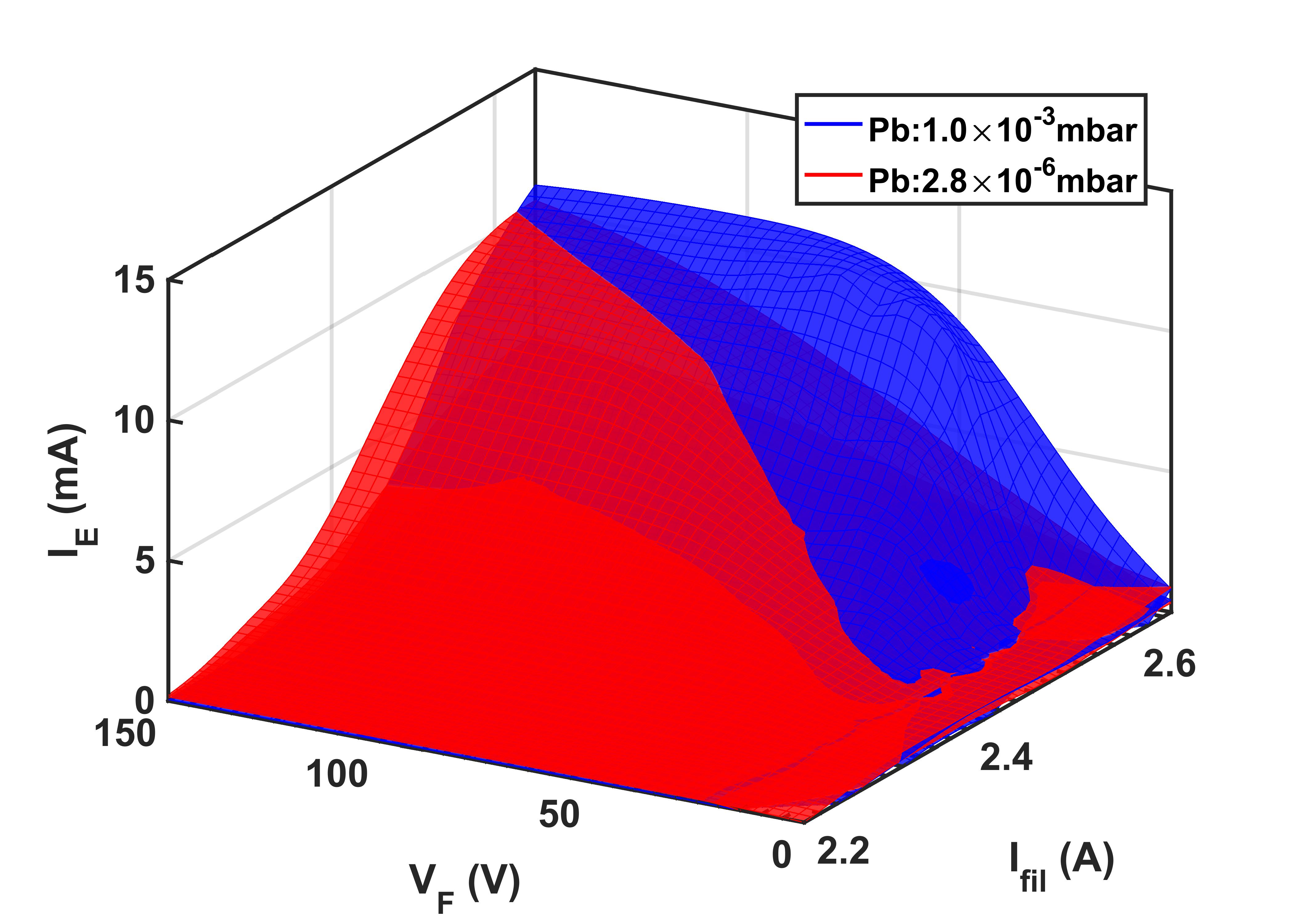}}
\subfigure[collection characteristics]
{\includegraphics[width=1\columnwidth, trim=0cm 0cm 0cm 0cm, clip=true,angle=0]{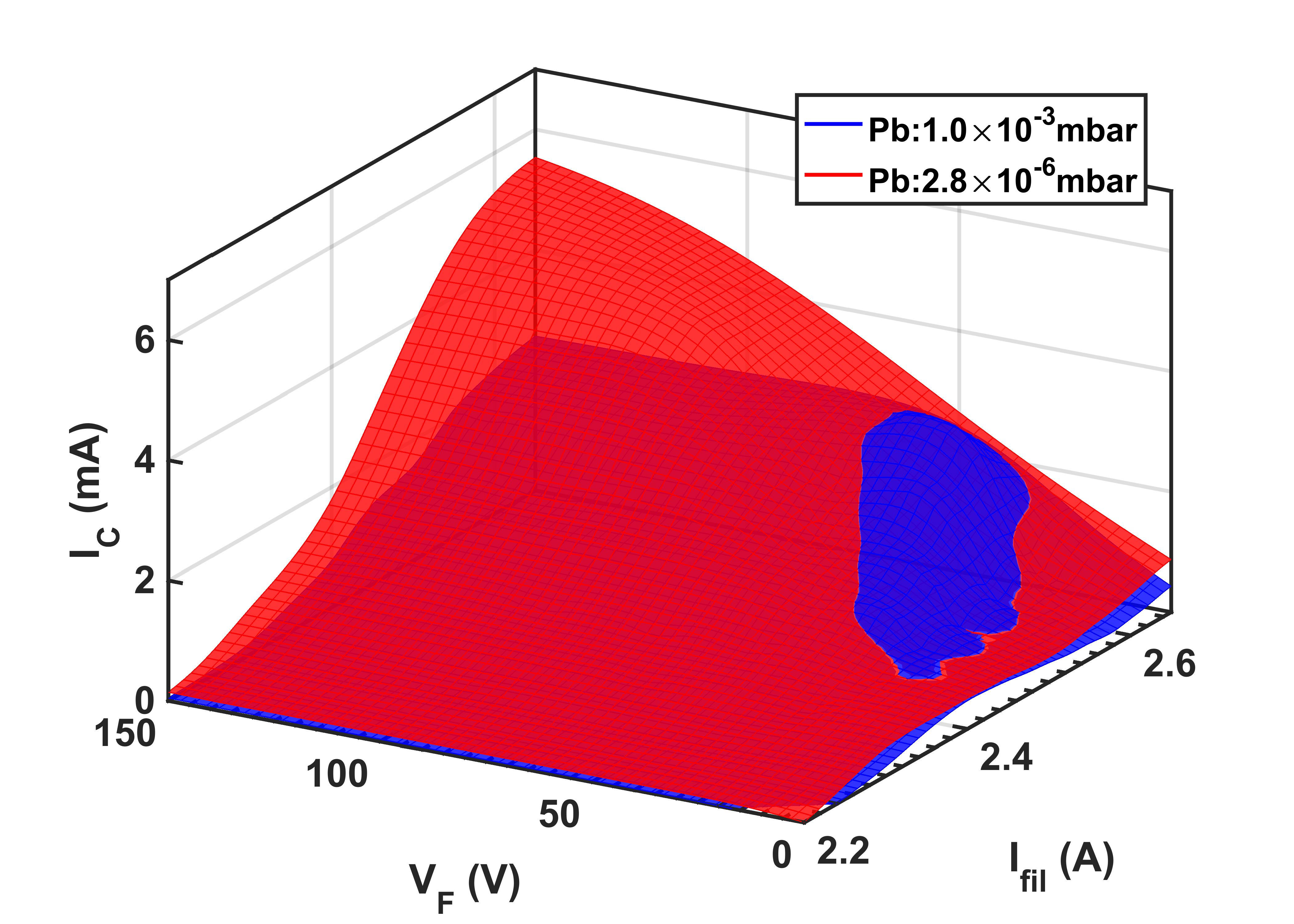}}
\caption{\label{fig:discharge_characteristics} Characteristics curves of SID-probe operating in discharge region (only fitting surfaces are shown for clarity)}
\end{figure}

Figure \ref{fig:discharge_characteristics} shows $I_E$ (a) and $I_C$ (b) characteristics curves with anode biased close to the ground potential (+10 V) at two difference background pressures. At lower pressure ($2.8\times10^{-6}$ mbar), discharge is absent shown in red (color online) and $I_E$ characteristics are same as that shown in figure \ref{fig:richardson_characteristics}, despite different biasing conditions. This again confirms that emission characteristics are solely determined by the relative biasing of the anode and cathode ($V_F$) and the functionality is not affected. Trends in the $I_C$ characteristics are also same as that of $I_E$ characteristics, but the absolute values are different, which is expected due to the loss of electrons. 
Shape of the characteristics surface remains unchanged with increase in the pressure, until discharge occurs, after which, it changes abruptly. Blue surface (color online) at background pressure $1\times10^{-3}$  mbar is the characteristics surface during discharge. Discharge forms a conductive path between the anode and cathode which reduces the space charge effects around the filament. As a result, filament emits more electrons, thereby increasing the $I_E$. Part of the discharge, which is attached to the vessel, provides an additional conductive path for the electrons to the vessel, which increases electron loss, thus decreasing $I_C$. During actual molecular beam measurements, discharge is limited to a small volume inside the probe, as discussed earlier in section-\ref{sec:con}. Hence, both $I_E$ and $I_C$ are expected to remain same. Once discharge occurs the shape of the characteristics surface remains unchanged, however, only the absolute value is affected.

\subsection{Optimizing the operating parameter of Probe}

The characteristics curves generated at different background pressures in Richardson and discharge regimes can be used to identify the combination of $V_F$ and $I_{fil}$ which can result in maximum signal during the beam measurements. For this we have to look into the discharge regime of the probe during the beam and during the characterization with background pressure. We define small localized discharge during the molecular beam measurement as the 'beam-discharge' and the discharge during the characterization using background pressure as the 'volume-discharge'. The characteristics of volume-discharge are already discussed as shown in figure \ref{fig:discharge_characteristics}. However, it is not possible to measure the characteristics of the beam-discharge. Hence, we assume same behaviour as in case of volume-discharge. As a result, beam-discharge will follow the trend of the volume-discharge in the localized region where the discharge is present.

The volume-discharge characteristic surface in figure \ref{fig:discharge_characteristics} shows that, for a particular filament current, the discharge shifts $I_E$ and $I_C$ saturation (knee region in Richardson regime) to lower field voltage $V_F$ (knee region in discharge regime). The saturation region is also the region where $I_C$ increases beyond its original value for Richardson regime which shows that, the signal will be maximum if the probe is operating in saturation region. Additionally, if the power supply of the probe is maintained at constant voltage (which is the case during beam measurement), a drop in filed potential $V_F$ is observed when discharge occurs. This is due to the increase in the conductivity of the discharge volume. Observed knee-shift and $V_F$-drop indicate that, signal strength will be maximum if $V_F$ prior to the discharge is such that, when discharge occurs, the probe is operating in the knee region (of discharge regime). As a result, it is essential to maintain $V_F$ higher so that it falls in the knee region when the discharge happens. Interestingly, our experiment shows that the fall in $V_F$ when discharge strikes is nearly same as the shift in knee region (between Richardson and discharge regime). For example, for $I_{fil} = 2.6 A$ , $V_F$ of 140 V at pressure $2.8\times10^{-6}$ mbar decreases to 50 V at $1\times10^{-3}$ mbar whre discharge is present. Hence, the shift in the knee-region in the characteristics surface can be used to identify the conditions to be maintained before the discharge so that $V_F$ falls to knee-region when discharge strikes.
 
As per the assumption stated earlier, during beam-discharge, the decrease in $V_F$ is expected to be the same as that in case of volume-discharge. However, due to the small size the beam-discharge, $V_F$ drop in localized region of the probe does not create significant overall drop for the entire electrode. To estimate the overall $V_F$ drop, we scale the localized $V_F$ drop to the ratio of 'Anode area for beam-discharge' to 'Anode area for volume-discharge' ($A_{orifice}/A_{internal surface}$). Assuming $V_F$ drop is same for both discharges, using this simple scaling, a $90V$ localized potential drop scales down to 77 mV. The experimental measured overall $V_F$ drop during 1.5 mS beam measurement is 120 mV. These is a reasonably good match considering the simplicity of the approach used. This validates the assumption of beam and volume discharges are nearly identical. Some variation due to geometry is expected ,however, they are difficult to quantify accurately. 
%As per our assumption, during beam-discharge, the decrease in $V_F$ is expected to be the same as that in case of volume-discharge. However, experimentally measured value is smaller for the beam-discharge (approximately 200 mV), because, $V_F$ drop in localized region of the probe does not create significant overall drop for the entire electrode. 

\begin{figure}[ht]
\centering
\includegraphics[width=1\columnwidth, trim=0cm 0cm 0cm 0cm, clip=true,angle=0]{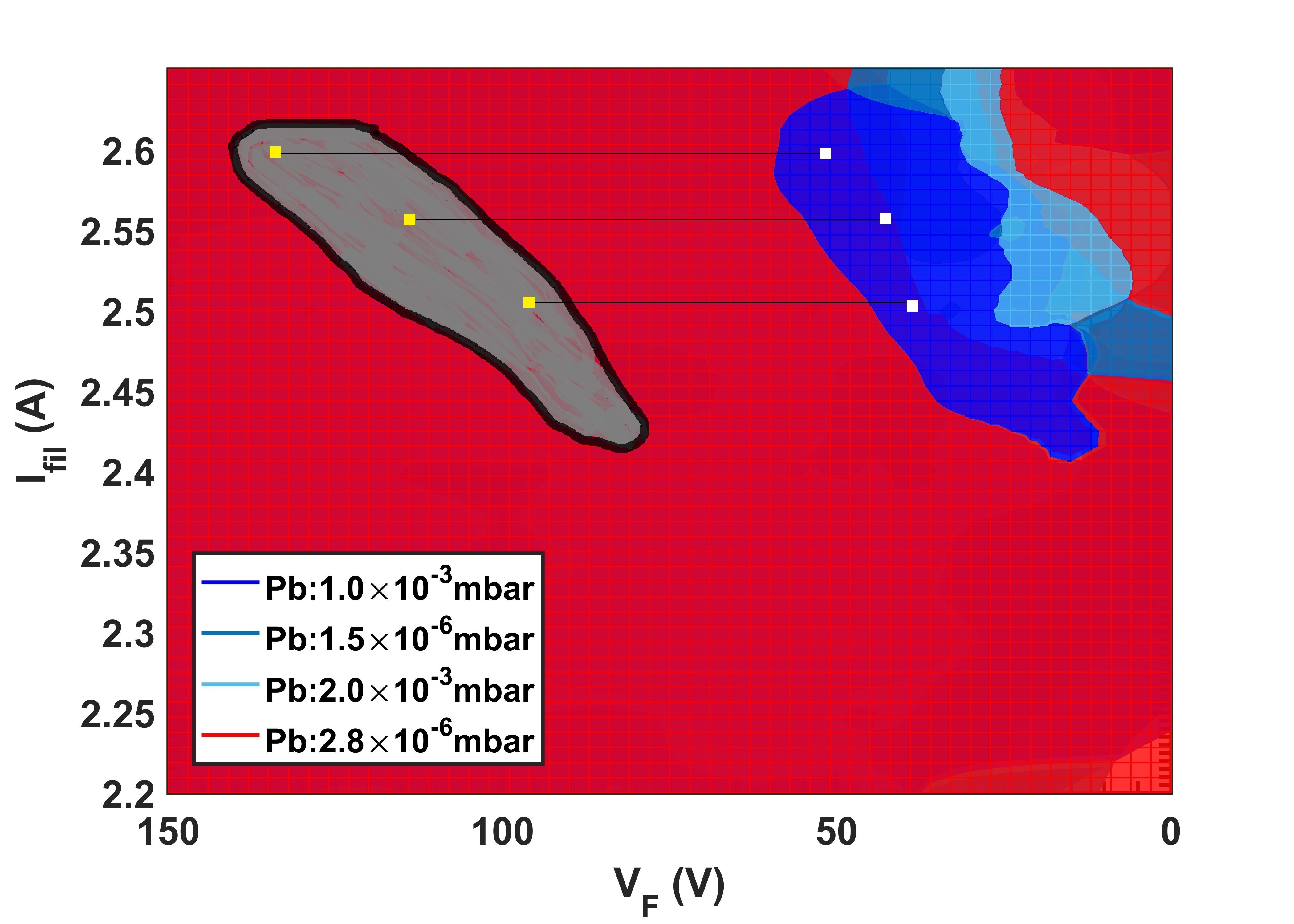}
\caption{\label{fig:operation_characteristics} $I_C$ characteristics at different pressures. Shades of blue indicates the region where $I_C$ is higher than original value in absence of the discharge (shown in red). Pairs of yellow-white square represents the drop in $V_F$ during discharge. Grey region shows the range of $V_F$ that gives best signal at a particular $I_{fil}$.}
\end{figure}

Figure-\ref{fig:operation_characteristics} shows reoriented $I_C$ characteristics surface of figure \ref{fig:discharge_characteristics}(b) before (Richardson regime, red) and after (discharge regime, shades of blue) the discharge for volume-discharge condition. It can be seen that the knee-region after the discharge (represented by different shades of blue) remains about the same for different pressures. Yellow and white squares for three individual values of $I_{fil}$ represent experimentally measured $V_F$ drop, before and after the volume-discharge at the constant power supply voltage. Hence, yellow squares indicates the conditions the probe needs to be for maximum signal. The grey area depicts the region where the signal is strongest during beam-discharge. This again shows that, the assumption of both the beam and volume discharges being identical is valid, and the characterization of volume-discharge gives a good approximation for optimum operating condition of the probe during beam-discharge. It is important to note that the grey region is the range of $V_F$, provides the best signal at a constant value of $I_{fil}$. Absolute value of the signal always increases with increase in both $I_{fil}$ and $V_F$. While it is desirable to find the optimal operational region for increased signal intensity, it is not mandatory for absolute density measurement. 

\section{calibration for absolute density measurement}\label{sec:cal}

As discussed in section \ref{sec:char}, both beam-discharge and volume-discharge are similar. As a result, the neutral density required for gas breakdown is also nearly same for both discharges. Thus, the number density during breakdown in volume-discharge is the same as the number density of the beam at the onset of the signal. Note that, the signal is $\Delta I_C$ which is the increment due to beam-discharge obtained over existing $I_C$ due to filament emission in absence of discharge. To measure the absolute density, for a particular operating condition ($I_{fil}$ and $V_F$), first the neutral density at the breakdown ($n_{brk}$) is calculated from the breakdown pressure in the volume-discharge. In the next step, under same operating condition, the vessel is evacuated and the molecular beam is injected. $\Delta I_C$ is measured for different beam density $n_{beam}$, which is generated by adjusting the nozzle pressure ($P_0$) of the source. This generates a plot of $\Delta I_C$ vs $P_0$. Beam density is a linear function of nozzle pressure $n_{beam} = kP_0$, where $k$ is the calibration factor. For, $n_{beam}<n_{brk}$, $\Delta I_C$ remains zero. After the breakdown ($n_{beam}>n_{brk}$), $\Delta I_C$ increases linearly with $P_0$. Transition in $\Delta I_C$ in indicates the breakdown condition. $P_0$ at transition can be measured experimentally by measuring $\Delta I_C$ after breakdown and extrapolating the trend towards lower $P_0$ to locate where $\Delta I_C$ becomes zero (intercept at $I_C=0$). Thus, from the $P_0$ at breakdown and $n_{brk}$ the calibration constant $k$ can be calculated. Using this calibration, $P_0$ can be converted to $n_{beam}$ to measure the absolute density. 

\begin{figure}[ht]
\centering
\subfigure[constant $I_{fil}$]
{\includegraphics[width=1\columnwidth, trim=0cm 0cm 0cm 0cm, clip=true,angle=0]{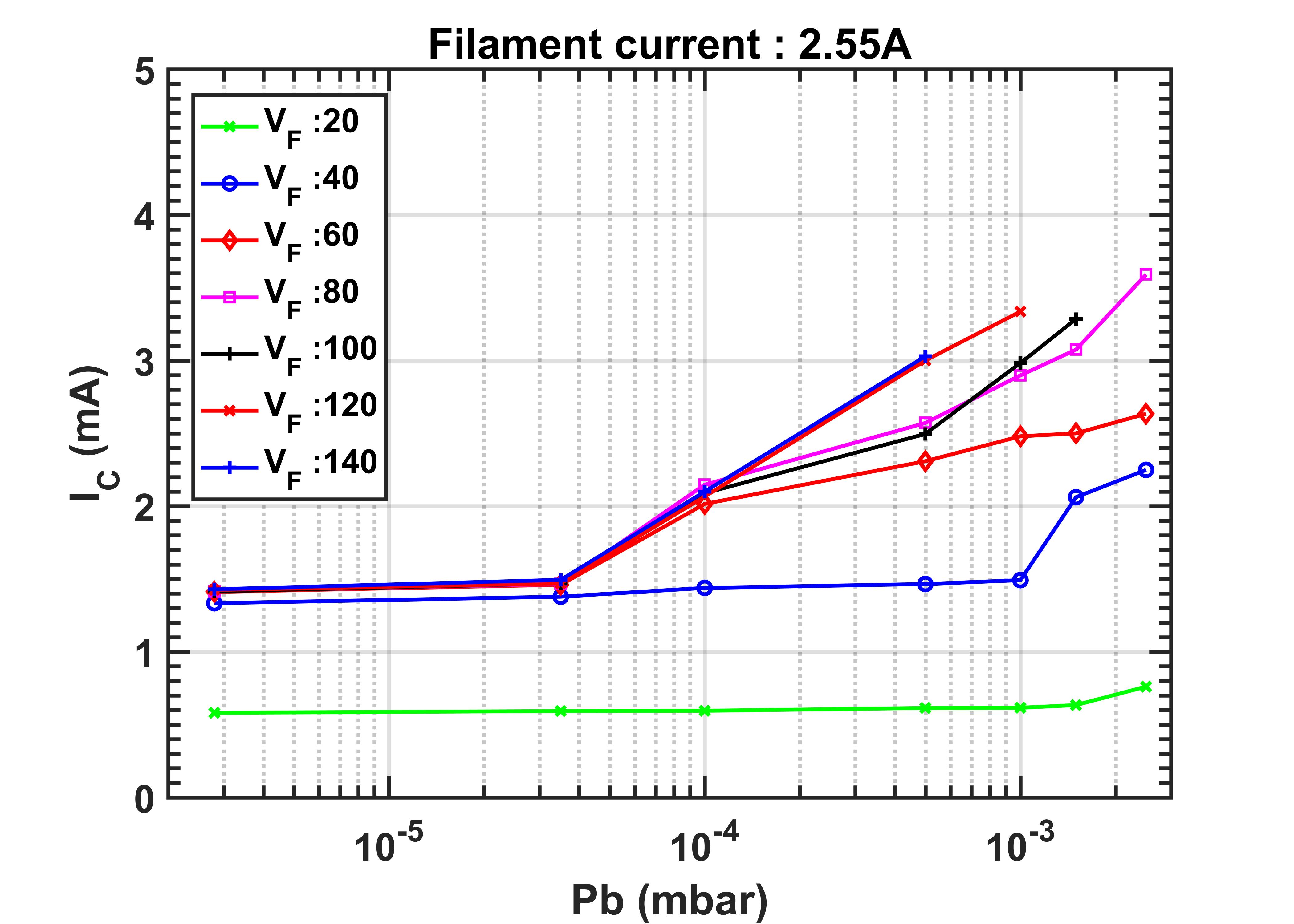}}
\subfigure[constant $V_F$]
{\includegraphics[width=1\columnwidth, trim=0cm 0cm 0cm 0cm, clip=true,angle=0]{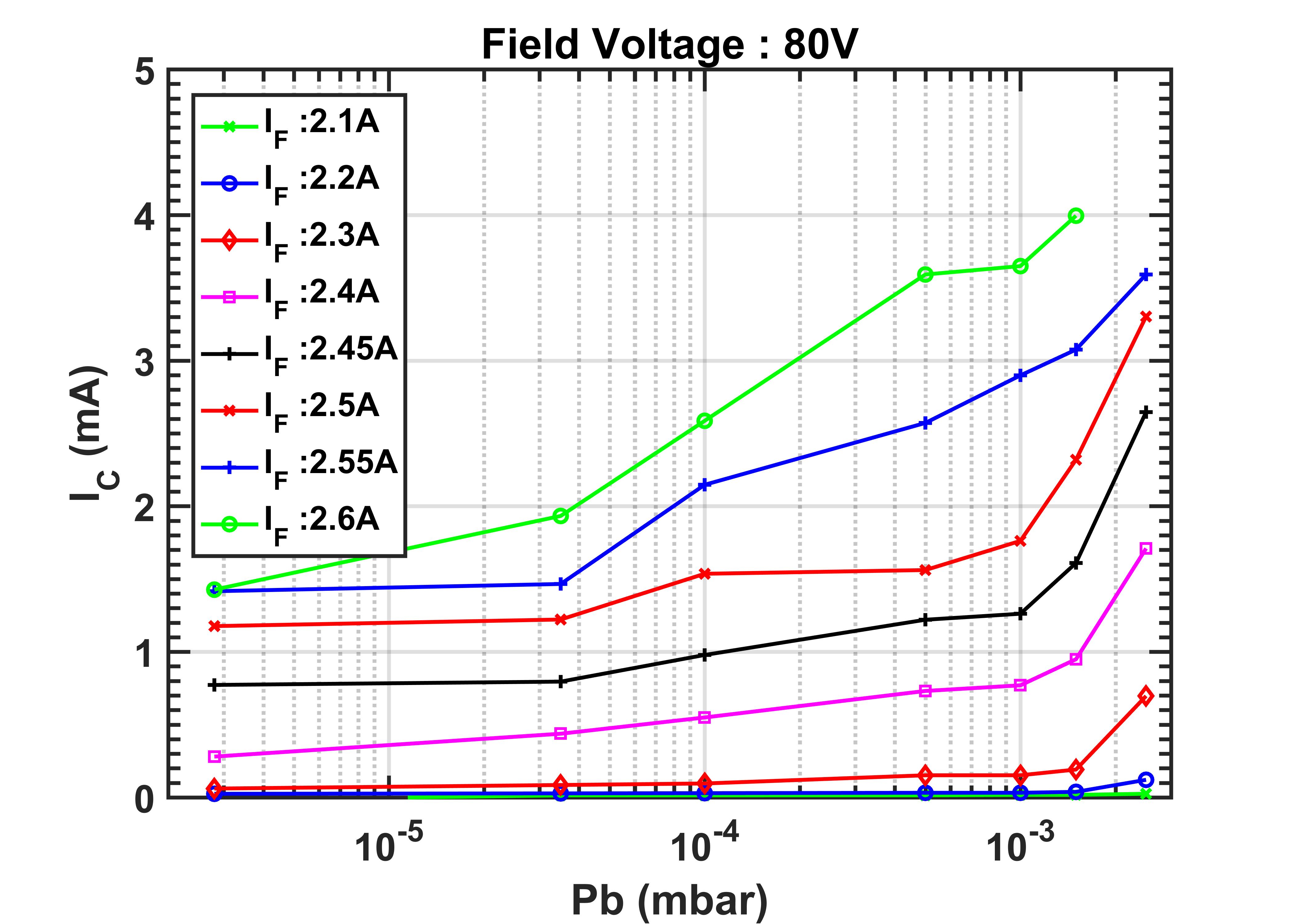}}
\caption{\label{fig:breakdown_characteristics} $I_C$ at different background pressures $P_b$ (a) different $V_F$ at constant $I_{fil}$, (b) different $I_{fil}$ at constant $V_F$. Increase in $I_C$ indicates the gas breakdown. $P_b$ limit saturates with $V_F$ but not with $I_{fil}$}
\end{figure}

%\begin{figure}[ht]
%\centering
%\includegraphics[width=1\columnwidth, trim=0cm 0cm 0cm 0cm, clip=true,angle=0]{breakdown_characteristics_const-I.jpeg}
%\caption{\label{fig:breakdown_characteristics_const-I} $I_C$ at different background pressures ($P_b$) for different $V_F$. Increase in $I_C$ indicates the gas breakdown. Breakdown pressure saturates for $V_F > 80V$.}
%\end{figure}
%
%
%\begin{figure}[ht]
%\centering
%\includegraphics[width=1\columnwidth, trim=0cm 0cm 0cm 0cm, clip=true,angle=0]{breakdown_characteristics_const-V.jpeg}
%\caption{\label{fig:breakdown_characteristics_const-V} $I_C$ at different background pressures ($P_b$) for different $I_{fil}$. At larger $I_{fil}$, breakdown condition shifts to lower pressure.}
%\end{figure}

The effect of heating current and field voltage on breakdown limit of volume-discharge is studied by monitoring $I_C$ at differing background pressure $P_b$, for different values of $I_{fil}$ and $V_F$. Figure-\ref{fig:breakdown_characteristics}(a) shows $I_C$ vs $P_b$ at different $V_F$ at constant $I_{fil} = 2.55A$ in volume-discharge condition, figure-\ref{fig:breakdown_characteristics}(b) shows it for different $I_{fil}$ at constant $V_F = 80V$. Rise in $I_C$ indicates breakdown. Figure \ref{fig:breakdown_characteristics}(a) shows that, at low field voltage, breakdown occurs at higher pressures. Increasing the field voltage allows breakdown to occur at lower pressures only up to certain limit, and further decrease is not observed for field voltage below 80 V. Figure \ref{fig:breakdown_characteristics}(b), on the other hand, shows that breakdown pressure limit does not saturate with the heating current. Increasing the heating current allows breakdown to occur at a lower pressure, thereby improving the density detection limit of the probe. As a result, in this work, the density measurements are recorded by operating the probe at relatively large $I_{fil} = 2.6 A$ and $V_F = 135 V$ during the calibration and measurements. Note that, \ref{fig:breakdown_characteristics}(b) also shows that $I_C$ increases by a small amount at lower pressure followed by a fast rise for higher pressures. This can be understood by the fact that breakdown occurs initially between the probe's anode and cathode, but, as pressure increases, discharge moves externally between the probe and the vessel leading to large rise in $I_C$. Since in molecular beam measurements, only internal discharge occurs, break down condition corresponding to internal discharge has to be taken into account to measure breakdown density. At operating condition of the probe ($I_{fil}=2.6A,V_F=135V$), the breakdown pressure was determined to be $3.5\times10^{-5}mbar$ and the corresponding number density is calculated using Loschsmidt number is $n_{brk}=8.56\times10^{17} molecules/m^3$. 

\begin{figure}[ht]
\centering
{\includegraphics[width=1\columnwidth, trim=0cm 0cm 0cm 0cm, clip=true,angle=0]{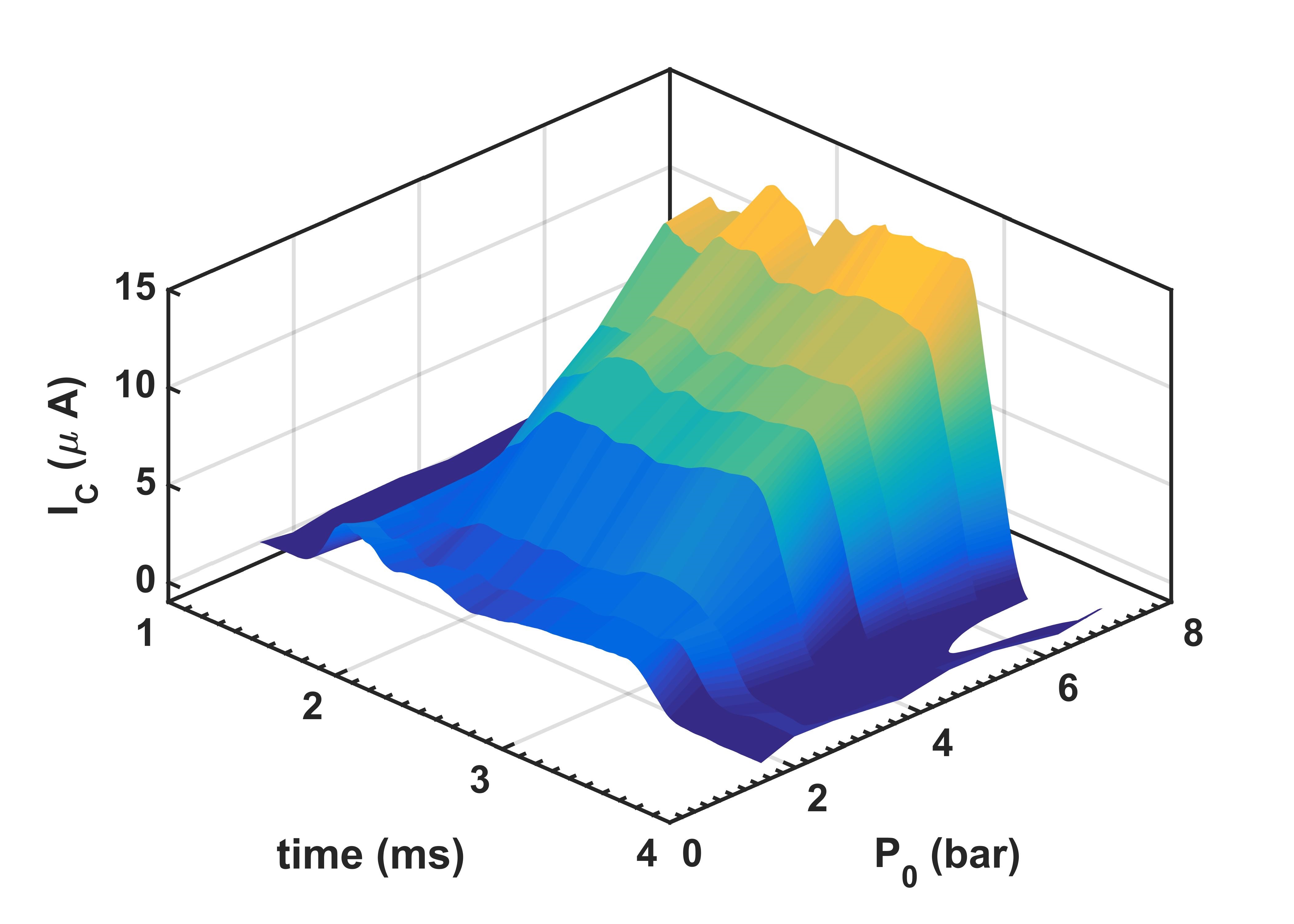}}
\caption{\label{fig:time_history} Instantaneous value of $I_C$ for different beam pressure achieved by changing reservoir pressure $P_0$.} 
\end{figure}

\begin{figure}[ht]
\centering
{\includegraphics[width=1\columnwidth, trim=0cm 0cm 0cm 0cm, clip=true,angle=0]{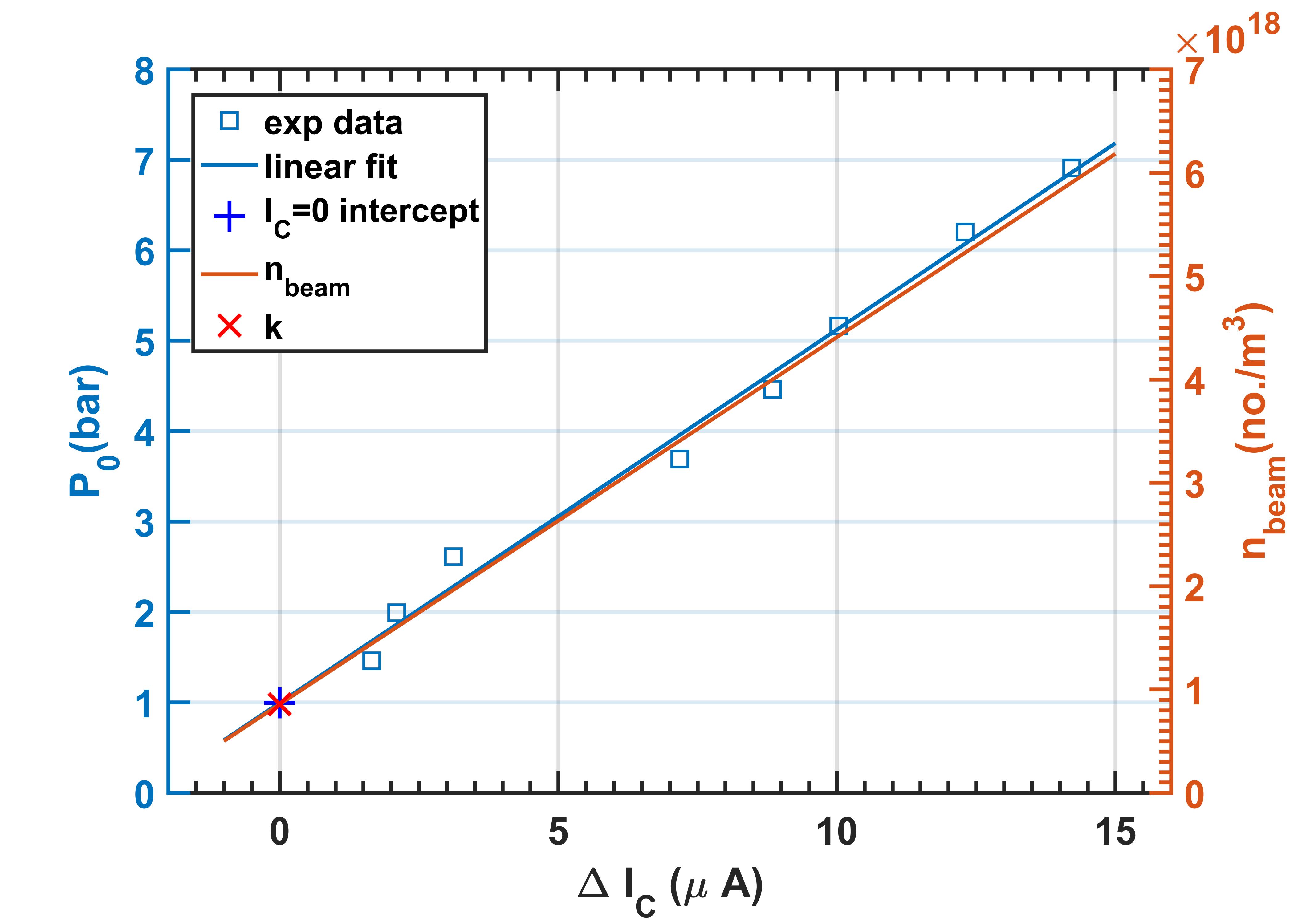}}
\caption{\label{fig:calibration} Calibration for absolute density. Time averaged values of $I_C$ (over plateau region) are plotted on X-axis. Y-intercept of linear fit (cross-marker) gives the $P_0$ at breakdown condition. Using calibration, $P_0$ is scaled in terms of number density of the beam ($n_{beam}$) shown on right Y-axis(red)}. 
\end{figure}

Figure \ref{fig:time_history} shows the time history of 1.5 ms molecular beam measured at the center of the beam for different nozzle pressures ($P_0$). Temporal width of $\Delta I_C$ indicates the beam duration, which is nearly equal to the 1.5 ms pulsewidth of the source. Presence of $\Delta I_C$ itself indicates that, the beam density is above the breakdown limit. As expected, $\Delta I_C$ increases linearly with nozzle pressure. Measurements below $P_0$ = 1.5 bar are not recorded as it was not possible in present experiment system. Flat plateau region in figure \ref{fig:time_history} indicates steady state of the beam. Instantaneous values of $\Delta I_C$ in steady state (plateau region) is averaged over time and are plotted on the X-axis in figure \ref{fig:calibration}. Corresponding $P_0$ is plotted on left y-axis. As discussed earlier, to determine $P_0$ at breakdown, a linear fit to experimental data is extrapolated at lower pressures where $\Delta I_C$ becomes zero. Intercept of the linear fit at $\Delta I_C=0$ gives $P_0=0.995$ bar at breakdown. From $n_{brk}=8.56\times10^{17} molecules/m^3$, the calibration factor $k=n_{beam}/P_0$ comes out to be $8.6072 \times 10^{17}$.
With this calibration factor, nozzle pressure $P_0$ is converted into beam density$n_{beam}$. These are plotted on the right y-axis (red) in figure-\ref{fig:calibration}. Factor $k$ is used to convert $\Delta I_C$ values to number density of the beam during the measurement of spatio-temporal profile of the beam.

\section{Measurement of spatio-temporal profile of beam}\label{sec:res}

Spatio-temporal profile of 1.5 ms molecular beam is measured by scanning the probe along the beam cross-section as discussed earlier. The density measured is averaged over the cross-section area of the hole in the shield that skims a fraction of the beam for measurement. Since, the hole diameter is 1.28 mm, scanning step size was selected to be 2 mm. A smaller step size will require deconvolution approach to calculate the absolute values, hence, was avoided. The selected step size, however, is sufficiently small to distinguish spatial variation of the density for anticipated beam size of 10 cm. At each location, measurements are averaged over 3 pulses. A quadratic surface is fitted over the recorded grid data to generate the beam profile.

\begin{figure}[ht]
\centering
\includegraphics[width=1\columnwidth, trim=0cm 0cm 0cm 0cm, clip=true,angle=0]{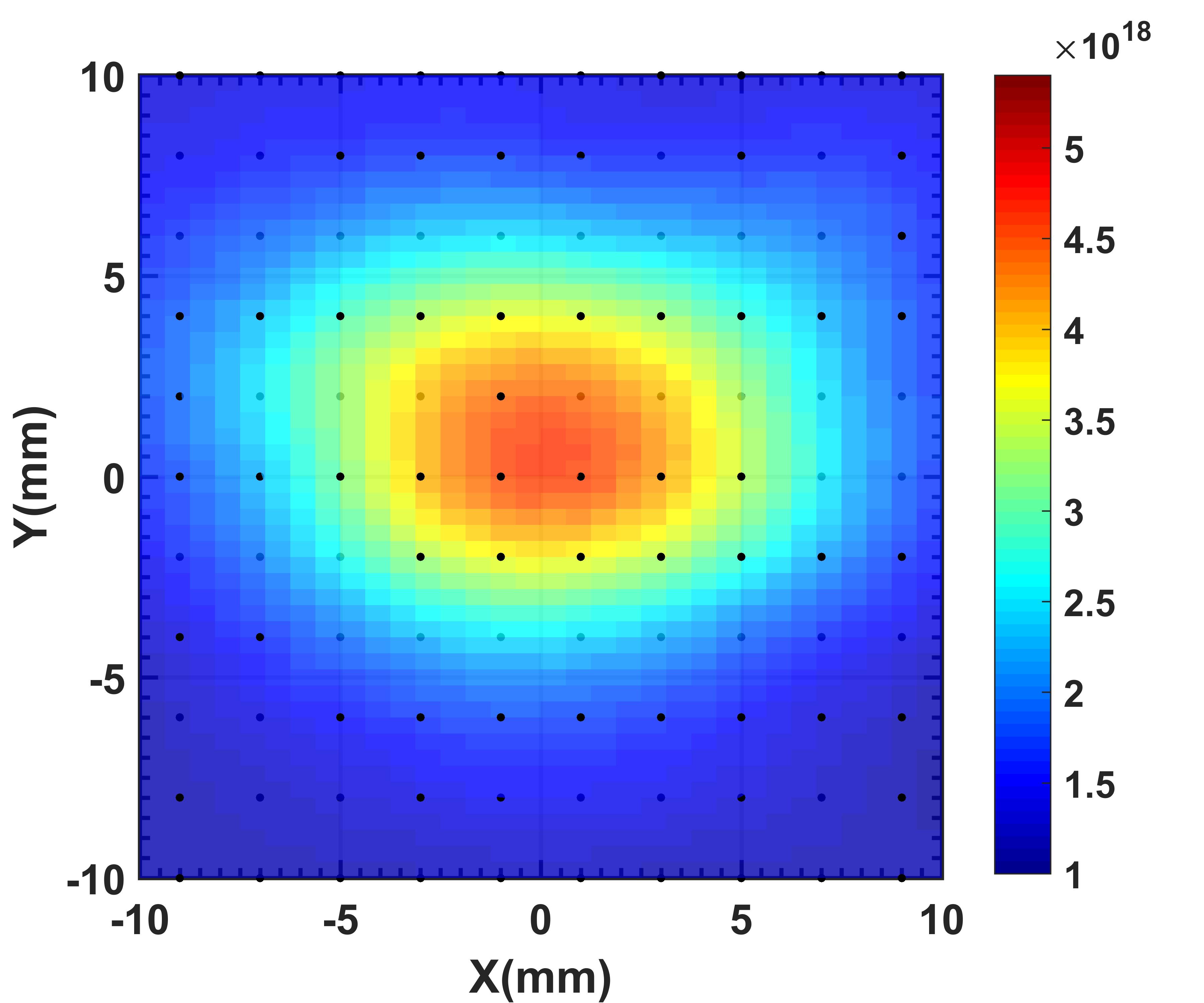}
\caption{\label{fig:time_avg_density} Profile of the 1.5 ms argon beam in time-averaged over stable flow period ($\backsim 1 ms$). Colorbar is scaled to absolute density ($molecules/m^2$). Dots represents the measurement location in experiments.}
\end{figure}

Figure \ref{fig:time_avg_density} shows the cross-sectional density profile of the 1.5 ms argon beam measured experimentally. Density values are time-averaged over stable flow duration of the beam (plateau region). A complete time history of the beam is provided as a video file in supplementary material (S1). The black dots represent the measurement location during experiments and the color plot representing density profile is a second order quadratic fitting surface with goodness of fit $R^2=0.97$. The beam profile is nearly circular with a minor Y-offset which is probably due to small vertical mis-alignment between nozzle and skimmer. Since the data are recorded for a $10\times10$ grid with an average of three pulses at each sampling location, the beam profile represents the combined data of 300 individual pulses. Despite large number of different individual pulses, no significant fluctuations were observed in measured values. This demonstrates the repeatability and measurement consistency of the probe. 

In order to evaluate the accuracy of the measurement, we calculated the expected beam density at probe location theoretically. In absence of skimmer, the axial number density at distance z from the sonic nozzle is given by model function \cite{bei1981}.

\begin{eqnarray}
n(z,\theta)=\frac{\kappa \dot{N}}{u_{\infty}\pi z^2} cos^b\left(\frac{\pi}{2}\frac{\theta}{\theta_{PM}}\right)
\label{eq:n_bjrnk}
\end{eqnarray}

An ideal skimmer only skims the para-axial region of the jet without disturbing it. Since, the skimmer in our experiment is located far from the nozzle in molecular flow region, we assume it as an ideal skimmer. However, in actual case slight skimmer interference may be present due to misalignment which reduces the actual density of the beam. As skimmer interference is difficult to accurately quantify, ideal skimmer assumption is a reasonably good approximation. Hence eq-\eqref{eq:n_bjrnk} holds for the skimmed part of the jet (molecular beam as it is known) and can be safely used to estimate the density. Centerline beam density at probe location (700 mm) calculated using eq-\eqref{eq:n_bjrnk} comes out to be $9.8\times10^{18} No./m^3$ while the density measured experimentally is $4.5\times10^{18} No./m^3$, which is an acceptable agreement considering the idealistic theoretical approach involved.

\section{conclusions}\label{sec:con}

This study conceptualizes a novel Shielded Ionization Discharge (SID) probe to measure the spatio-temporal profile of a pulsed supersonic molecular beam and the working is demonstrated by measuring the spatio-temporal density profile of 1.5 ms Argon beam. 
The probe operates using small discharges which are assisted by thermionic electron emission from a hot filament. The probe is characterised in Richardson and discharge regions and the ideal operating range is determined by emission and collection characteristics surfaces. Calibration for absolute density is carried out by comparing the breakdown condition at known background pressure with the signal trend for molecular beam measurement at different nozzle pressures. Briefly, SID probe is used to generate a complete spatio temporal density profile of molecular beam. Experimentally measured time-averaged centreline density is in fair agreement with the theoretical predictions considering ideal approach involved. We believe that the proposed SID probe can be used to measure and optimize molecular beams used in many applications.

\section*{S1. Supplementary material}

Video on time history of density profile of 1.5 ms molecular beam.

\section*{Data availability} 

The data that supports the findings of this study are available from the corresponding author upon reasonable request.

\section*{references}

\end{document}